# Electron temperature dependence of the electron-phonon coupling strength in double-wall carbon nanotubes

Ioannis Chatzakis[1,a),b)]

[1] Department of Physics, JRM Laboratory, Kansas State University, Manhattan, KS 66506, USA

We applied Time-Resolved Two-Photon Photoemission spectroscopy to probe the electron-phonon (*e-ph*) coupling strength in double-wall carbon nanotubes. The *e-ph* energy transfer rate $G(T_e,T_l)$ from the electronic system to the lattice depends linearly on the electron ($T_e$) and lattice ($T_l$) temperatures for $T_e > \Theta_{Debye}$. Moreover, we numerically solved the Two-Temperature Model. We found: (i) a $T_e$ decay with a 3.5 ps time constant and no significant change in $T_l$; (ii) an *e-ph* coupling factor of $2 \times 10^{16}$ W/m$^3$; (iii) a mass-enhancement parameter, λ, of $(5.4 \pm 0.9) \times 10^{-4}$; and (iv) a decay time of the electron energy density to the lattice of 1.34±0.85 ps.

[a)] Electronic mail: ichatzak@slac.stanford.edu

[b)] Present address: Department of Materials Science and Engineering, Stanford University, Stanford, California 94305, USA and Stanford Institute for Materials and Energy Sciences, SLAC National Accelerator Laboratory, Menlo Park, CA 94025, USA

In carbon nanotubes[1] and graphene[2], the electron-phonon (*e-ph*) interactions modify the dynamics of the charge carriers near the Fermi level by changing their mass and the relaxation rate. A dramatic manifestation of these interactions is found in superconductivity, in ballistic transport[3-9] phenomena in Raman spectra,[10] and in phonon dispersion.[11] A significant rise in the electron temperature, $T_e$, with respect to the lattice temperature, $T_l$, can be achieved by irradiating the material with ultra-short laser pulses. The *e* and *ph* systems equilibrate by exchanging energy (via scattering processes) with a rate defined by the coupling strength. The observed energy bottleneck suggests that only a limited set of the total phonon modes[12, 13] participate in the relaxation of the charge carrier energy. The *e-ph* coupling strength is highly important, and although it has been extensively investigated in single-wall carbon nanotubes (SWNTs)[4, 9, 14, 15] and graphene,[16] very little is known for double-wall nanotubes (DWNTs).

In this letter, we present the temperature dependence of the energy transfer rate $G(T_e,T_l)$ from electrons to the lattice in DWNTs. Using time-resolved two-photon photoemission (TR-TPP) spectroscopy[17], we record the non-thermal and the thermal evolution of the electron distributions and their subsequent equilibration with phonons. We analyze the dynamics of the excited electrons in the vicinity of the Fermi level and determine the *e-ph* interactions. We also solve numerically the Two-Temperature Model (TTM), which describes the hot electron and phonon energy evolution after the excitation. The *e-ph* coupling arises from the coupling in the inner tube, the coupling in the outer tube, and the coupling from electrons scattered from the inner to the outer tubes, though no significant increase in the total *e-ph* coupling in DWNTs compared to separate SWNTs was found[18].

In our experiment, IR ultra-short laser pulses were utilized to generate a non-equilibrium population of charge carriers in the conduction band by absorbing energy from the pulses.

Subsequently, the excited electrons, ionized by the UV probe pulses, were delayed with respect to the excitation pulse by a mechanical variable delay stage. Recording the time-of-flight of the ionized electrons monitors the dynamics of the hot electrons in the conduction band. The pump beam consisted of IR laser pulses of ~35 fs duration at an 1.55 eV photon energy provided by a Ti:Sapphire laser amplifier with an 1 KHz repetition rate. The pump beam with fluence ~40 µJ/cm$^2$ was linearly S-polarized. The P-polarized probe pulses of 4.71 eV photon energy (having fluence ~30 nJ/cm$^2$) were obtained by frequency tripling of the fundamental pump pulse through non-linear effects during the photoionization of N$_2$ molecules. The fluence of the pump beam was kept relatively low in order to avoid space charge effects (<50 e/pulse), multi-photon processes during the excitation, and transient thermal heating of the lattice. Thus, linear absorption of light occurred. Photoelectron spectra were obtained by the time-of-flight technique using a spectrometer with 10 meV energy resolution, at 1 eV electron kinetic energy ($E_{kin}$). Further details of the experimental setup can be found elsewhere.[19] Our sample was 100 µm thick freestanding ''bucky'' paper with an outer diameter of the nanotubes of 4±1 nm provided by the Nanolab.

The probe-photon energy exceeds the work function of the sample by ~0.25 eV, thus directly promoting electrons from below the Fermi level, $E_F$, to the vacuum level. The initial state energy ($E - E_F$) of the photoemitted electrons with respect to $E_F$ is obtained from their $E_{kin}$, since ($E - E_F$) = $E_{kin}$ + $e\Phi$ − $h\nu_{probe}$, with $e\Phi$ the work function of the sample and $h\nu_{probe}$ = 4.71 eV the energy of the probe beam. To isolate the photoelectrons originating from the excited states above $E_F$, the spectrum due to the probe pulse only was subtracted from the spectrum due to both pump and probe pulses. The difference ($\Delta I = I_{pump+probe} - I_{probe}$) between the two spectra is illustrated in Fig. 1 for several time delays. This ''excitation difference'' reflects the non-equilibrium carrier distribution as induced by the pump pulse. The negative signal below $E_F$ is due to vacancies induced by the pump pulse (decrease of the electron densities below $E_F$). In contrast, the population of electrons above the $E_F$ increases (positive signal) due to the migration of electrons originating from states below $E_F$. The maximum amplitude of these difference spectra occurs at ~100 meV above and below $E_F$ for electrons and holes, respectively. The amplitude decreases as the delay increases, i.e. the energy stored in the electronic system decays out of the excited volume to the bulk. A fit to the Fermi-Dirac (F-D) distribution of the experimental data is usually used to determine $T_e$ and to estimate the internal energy of the electronic system. Thus, we used the difference at a certain delay between the F-D functions $f(T_e)$, which represents the pump-induced change in the electron distribution, and $f(T_l)$. Figure 2a shows two calculated Fermi distributions at 300 K and 2000 K and Fig. 2b their difference, which fits the experimental data and specifies $T_e$ (with free parameters $T_e$, and a proportionality constant). The excited electron distribution deviates from the Fermi statistics for short time delays between the pump and probe pulses. This is caused by the finite time required by the system for reaching its internal equilibrium through $e$-$e$ (Coulomb) interactions after the perturbation by the pump pulse. Due to the conical energy spectrum (Dirac cones) at the K (K') points of the Brillouin zone, electronic states occur only close to K (K'). The available phase space for the Coulomb interactions becomes vanishingly small as $E$ approaches $E_F$, particularly after the initial avalanche of the hot electrons towards $E_F$. A bottleneck can be created resulting in a lowering of the internal thermalization rate. After a few hundred femtoseconds, the internal thermalization has been established. Hence, the electron distribution can be described by a F-D statistics with a definable temperature, $T_e$. In multi-wall carbon nanotubes and graphite, the internal thermalization occurs at ~0.2 ps,[19, 20] but for metallic systems it is significantly faster (~10 fs[21]). The (Fig.2b) shows the data for the electron distribution at 570 fs after the excitation, well described by the F-D distribution with an elevated temperature of 2000±100 K. The cutoff for the spectrum is at ~1.5 eV (determined by the pump photon energy), while only a limited number of carriers are excited up to this energy.

A commonly accepted theory for the energy relaxation dynamics between electrons and lattice is the TTM, which assumes that the electron and phonon subsystems are each maintained in a thermalized state (local equilibrium) by Coulomb (*e-e*) and unharmonic (*ph-ph*) interactions[22, 23], respectively. The $G(T_e, T_l)$, which occurs when $T_e > T_l$, can be obtained by probing the dynamics of the excited electrons above the Fermi level, following excitation by a femtosecond laser pulse. It depends upon the coupling strength between the two systems, which, in turn, is directly related to the mass-enhancement parameter $\lambda$[21, 24] that is of crucial importance in the BCS theory.

The time evolution of the energy in the two subsystems is described by[25]

$$C_e(T_e)\frac{dT_e}{dt} = \nabla(k\nabla T_e) - G(T_e, T_l) + S(x,t) \qquad (1)$$

$$C_l(T_l)\frac{dT_l}{dt} = G(T_e, T_l), \qquad (2)$$

where $C_e$ (J/m³·K) is the electronic heat capacity, $\kappa$ (W/m·K) is the thermal conductivity, $G(T_e,T_l)$ (W/m³·K) is the coupling term, $S(t)$ is the source term for the absorbed laser energy density per unit time, and $C_l$ (J/m³·K) is the lattice heat capacity with $C_l \gg C_e$. The term $\nabla(\kappa\nabla T_e)$ describes the diffusive electronic heat transport out of the excited region and has been omitted, since this process is too slow relatively to the timescale of the observed electron-cooling rate. In DWNTs, $C_e \approx C_{oph}/100$ for the optical phonons[26] that are strongly coupled with the electrons, while $C_{oph}=2.3\times10^6$ J/m$^3$·K at 300 K[27], which is almost twice that in SWNTs ($1.37\times10^6$ J/m$^3$·K)[28]. At room temperature, $C_e$ is much smaller (~$1.51\times10^4$ J/m$^3$·K). Thus, the phonon temperature does not increase noticeably after the *e-ph* equilibrium. We also assumed linear absorption of light, i.e. the light intensity decreases exponentially with depth in the material with a penetration length δ~17 nm at 800nm. The system of the two coupled equations (Eqs.1,2) was solved numerically to predict the time dependence of $T_e$ and $T_l$.

The electron gas right after the thermalization, due to *e-e* interactions is far away from equilibrium with the lattice, even though it has acquired internal equilibrium. At this stage, the electrons follow the F-D distribution with characteristic temperature, $T_e$, which represents the excess energy of the non-equilibrium electrons (initial free-carrier excess energy) and corresponds to the electron energy measured above the conduction band minimum and the holes energy measured below the valence band maximum.

Using the TR-TPP spectroscopy, we determined experimentally the $G(T_e, T_l)$ for $T_e > \Theta_{Debye}$, which can be written in terms of the internal energy, $E_{int}$, of the electronic system, $G(T_e,T_l) = -dE_{int}/dt$. The $E_{int}$ (J/m$^3$) is calculated using the relation, $E_{int} = \gamma T_e^2$, where $\gamma \sim 44$ J/m$^3$K$^2$ is the electronic heat capacity coefficient (in agreement with Shi *et al.*) and $T_e$ is derived from the F-D fit of the photoexcited electron distribution above $E_F$ (Fig. 2b). Figure 2c shows the temporal evolution of the internal energy of the electronic system. The transient decay time is determined by fitting the experimental data with a single exponential decay with a time constant τ. We found τ =1.34±0.85 ps at room temperature for the relaxation of the electron energy. The absorbed energy from the pump pulse causes $T_e$ to increase to ~2350 K, while the lattice remains in equilibrium. The photoexcited electrons transfer most of their energy to the more strongly coupled optical phonons (SCOPs),[12] including the zone-center (G-mode) and K-point phonons with a characteristic time of a few hundred femtoseconds and effective temperature $T_{oph} \approx T_e$ (with $T_{oph}$ the temperature of SCOPs). In turn, this non-equilibrium hot phonon population dissipates its energy into the secondary lower-energy phonons[22, 23] (acoustic modes) through anharmonic interactions and a time constant of 1-2 ps at room temperature, as has been reported previously by Chatzakis *et al.*[22] and others[22, 29-31] for SWNTs and graphite. The energy loss for the acoustic phonons to the substrate takes place in tens of picoseconds.[32] The corresponding heat capacities are $C_{oph}$ for the optical phonons and $C_{ac}$ for the acoustic phonon bath with $C_{ac} \gg C_{oph}$ thus, the change in $T_l$ is less than the temperature of SCOPs (Fig. 3). We note that the dynamics of the hot electrons and the SCOPs become similar after a few hundred femtoseconds. The time constant for the energy decay here (1.34 ± 0.85 ps at room temperature) is quite similar to that found by Chatzakis *et al.*[22] (0.6 – 1.2ps for high and low temperatures, respectively) in SWNTs.

The derived temporal evolution of $T_e$ and $T_l$ as described by the TTM is presented in Fig. 3 along with the experimental data for comparison. The best agreement with the measured data is achieved for a *G(T)* coupling factor of $2 \times 10^{16}$ W/m³ at room temperature, with typical values for noble metals (e.g., Au) of $2.3 \times 10^{16}$ W/m³. The equilibration between $T_e$ and $T_l$ is completed at ~6 ps, corresponding to a $T_l$ increase by only ~60 K. Because $C_{ac} \gg C_{oph}$, after ~4 ps $T_l$ is stabilized at 360 K for several tens of picoseconds, even though $T_e$ has not relaxed complete. The electron temperature $T_e$ acquires a local equilibrium having reached the same temperature with the lattice, $T_l$~360 K, after ~6 ps. The decay time constant for $T_e$ is ~3.5 ps at 1/e of the temperature amplitude, as it can be estimated from the calculated transient temperature in Fig. 3, in very good agreement with the results for graphene[12, 33-35], for which the equilibration of $T_e$ with the optical phonons is completed in 3.1 ps. However, the equilibration time in DWNTs can differ for different nanotube species, since the *e-ph* coupling strength depends on the density of states and the transition matrix elements, which are both affected by the curvature difference between the inner and the outer tubes. Due to the nonlinear relation between the energy density of the electrons, $E_e$, and $T_e$, $\left(T_e \propto \sqrt{E_e}\right)$, their temporal evolution is different.

The $G(T_e, T_l)$ is obtained by the differentiation of the experimental data of Fig. 2c with respect to time and plotted as a function of the temperature difference, $\Delta T=(T_e - T_l)$, where $T_l =$ 300K $\approx$ constant, since the rise of $T_l$ after the equilibration is only ~60 K. For the low temperature limit ($T_e$, $T_l << \Theta_{Debye} \approx$ 2000 K), $G(T_e, T_l)$ depends upon the 5$^{th}$ power of $T_e$ and $T_l$ [36], while for $T_e$, $T_l > \Theta_{Debye}$, $G(T)$ is linear with T.[21] In this work, the high temperature data ($T_e$ of the order of $\Theta_{Debye}$) are fitted with a model that is linear with the temperature by using $G(T) = g(T_e - T_l)$, as shown in Fig. 4, with the slope, $g$, of the line as a free parameter, and $T_l$ fixed at 300 K. From the fit, a slope, $g$, of $(6 \pm 1) \times 10^{15}$ W/m$^3$K was obtained. For comparison, $G$ is also shown as a function of the 5$^{th}$ power of the temperature, which deviates from the data. Allen's theory relates $g$ to the second moment, $\lambda<\omega^2>$, of the Eliashberg's spectral function as $g = 3\hbar\gamma\lambda<\omega^2>/\pi k_B$. Using the latter relation, the approximation $<\omega^2> \approx (\Theta_{Debye})^2/2$ with $\Theta_{Debye} =$ 2000 K, and $\gamma =$ 44 J/m$^3$K$^2$, we estimate the second moment, $\lambda<\omega^2>=8\times10^{-3}$ (meV)$^2$, and the mass-enhancement parameter, $\lambda = (5.4 \pm 0.9)\times10^{-4}$, which is smaller than that in SWNTs[37]. Comparing $G(T)$ for SWNTs provided in Fig. 19 of Ref. 37 and DWNTs (i.e., at 1000 K), one can see that $G(T)$ for SWNTs is larger by an order of magnitude. In principle, this implies that a higher density-of-states $N_F$ near the Fermi level contributes in SWNTs than in DWNTs. Here, this is not the case, since $N_F$ is significantly higher for DWNTs than SWNTs, e.g., $N_F$ for the metallic species (5,5) and (10,10) is equal to 0.607 states/eV/spin/cell and 0.570 states/eV/spin/cell, respectively, while for DWNTs (5,5)@(10,10) is 1.190 states/eV/spin/cell.[18]

Thus, λ should be smaller in DWNTs. Indeed, λ was found to be $(5.4\pm0.9)\times10^{-4}$, i.e. smaller by an order of magnitude than that for SWNTs[38]. In general, the coupling strength varies from one phonon band to another; it decreases when the tube diameter increases, but the variation on the coupling strength with the diameter differs from one mode to another.[39] In our experiment, the diameter of the measured tubes is of the order of ~4 nm, significantly larger than that used by Connetable *et al*.[39] for SWNTs. Here, we mostly measure $G(T)$ in metallic nanotubes, due to the higher $N_F$ between the Van-Hove singularities from case to case. The DWNTs show peculiar electronic structures due to the interactions between the inner and the outer tubes and the curvature effects, resulting in a local density-of-states in the gap. These two effects are also responsible for the transition from semiconductor to metal, e.g., the energy gap in the DWNT (7,0)@(16,0) vanishes due to the merging of the π and π* states, although each constituent nanotube is a semiconductor with finite energy gap. Similarly, in the (7,0)@(17,0) the π and π* states overlap resulting in the metallization of the DWNT[40].

Summarizing, we applied TR-TPP spectroscopy in DWNTs to probe the charge carrier dynamics near the Fermi level. After excitation, the electrons are internally thermalized through Coulomb interactions and follow F-D statistics with a characteristic temperature $T_e$. The electrons are equilibrated with the SCOPs within a few hundred femtoseconds and their temperatures become similar. These SCOPs provide an efficient channel for the electron energy equilibration. We measured a time decay constant for the energy of the hot electrons of 1.34 ± 0.85 ps, in agreement with previous studies. Furthermore, we performed theoretical simulations based on the TTM, (see Supplementary Material) [41] and obtained quantitative information about the evolution of $T_e$ and $T_l$. At room temperature, we estimated an *e-ph* coupling factor of $2\times10^{16}$ W/m$^3$. The simulations showed that $T_e$ decays with a time constant of 3.5 ps, in very good agreement with the results for graphene[33]. For $T_e \leq \Theta_{Debye} \approx 2000$ K and using Allen's theory, we found a linear temperature dependence of $G(T_e,T_l)$ and a mass-enhancement parameter, $\lambda$, of $(5.4\pm0.9)\times10^{-4}$, suggesting ballistic conductance of the carriers in DWNTs in this regime.

## Acknowledgements

This work was supported by the Chemical Sciences, Geosciences, and Biosciences Division, Office of Basic Energy Sciences, Office of Science, US Department of Energy. The author gratefully thanks Prof. E. N. Economou for useful discussions and acknowledges Prof. P. Richard and Dr. P. Tassin for helpful suggestions; these contributions have helped the quality of this paper.

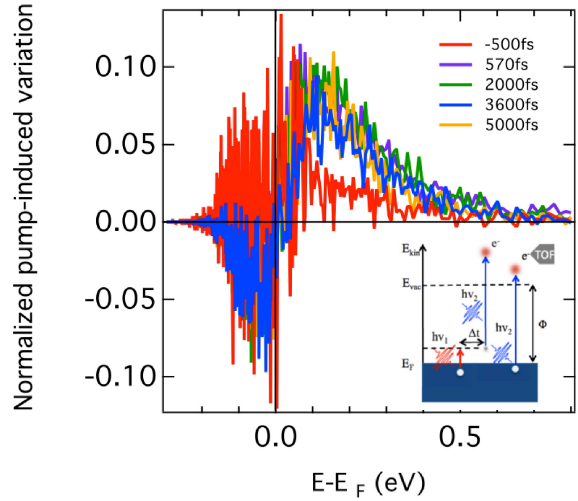

**Fig. 1** Time evolution of the photo-excitation spectra for different time delays showing the changes in the electron distribution due to the absorbed energy from the IR pump pulse. (*Inset*) Photo-excitation of the electrons from below the Fermi level to the continuum.

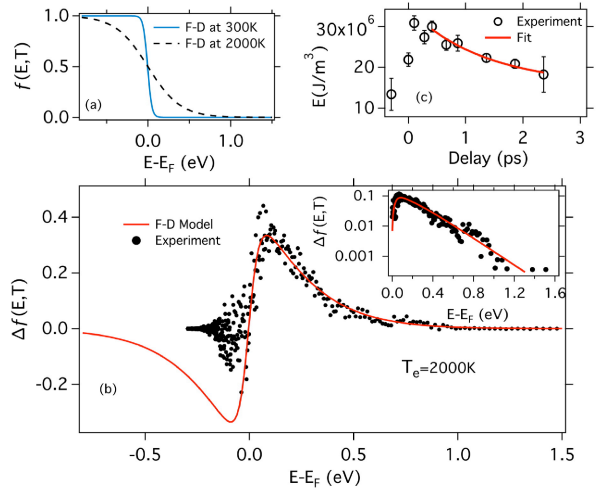

**Fig. 2** (*a*) F-D distribution for two different electron temperatures, $T_e$, 300 K and 2000 K. (*b*) Electron distribution above the Fermi level probed at 570 fs after the excitation and modeled as the difference $\Delta f(E,T)$ of two F-D distributions (*inset* in log scale) with an elevated $T_e$ of 2000±100 K and 300 K. (*c*) Temporal evolution of the electron energy density $E_{int}(T_e)$ fitted with an exponent with a time constant $\tau$ = 1.34 ± 0.85 ps, restricted at $\tau$ > 500 fs delays, where the internal thermalization of the electronic system is established. The error bars represent the uncertainty in the determination of the internal energy of the electronic system calculated using the temperatures obtained from the F-D fit of the excitation spectrum.

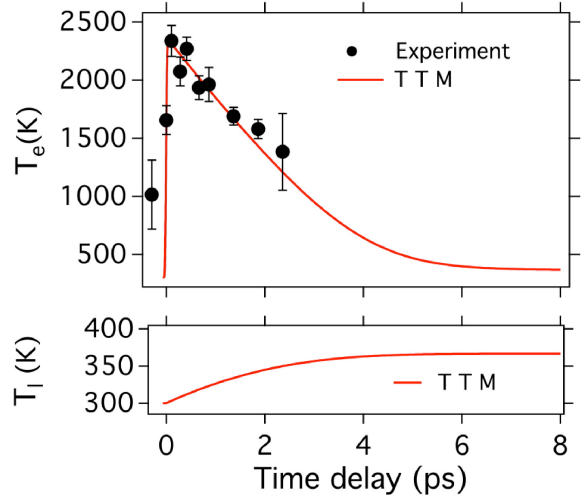

**Fig. 3** Temporal evolution of the electron, $T_e$, and lattice, $T_l$, temperatures based on the TTM (red lines). The measured cooling rate of the electronic system is shown with black dots. The error bars represent the errors of the fit of the F-D function to the photo-excited electron distribution above $E_F$.

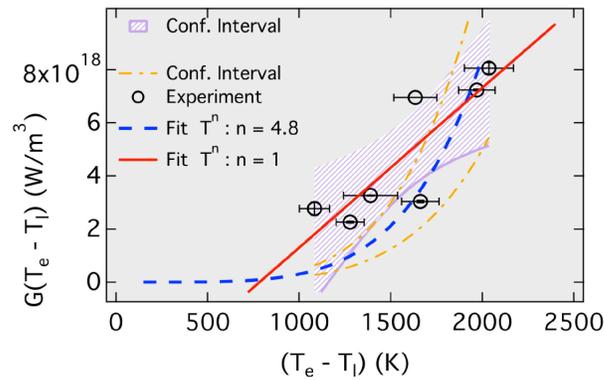

**Fig. 4** Energy transfer rate $G(T_e,T_l)$ from electrons to lattice as a function of their temperature difference fitted to a linear relation of $T$ (solid red line; 95% confidence interval as purple shaded area) with the slope $g$ as an adjustable parameter. The vertical error bars are related to the error of the fitted $G(T_e,T_l)$, while the horizontal error bars are $T_e$ estimations. The blue dashed line shows the dependence on the fifth power of the temperature (95% confidence intervals in yellow dot-dashed lines).

# SUPPLEMENTARY MATERIAL

**Electron temperature dependence of the electron-phonon coupling strength in double-wall carbon nanotubes**

*Ioannis Chatzakis*[a),b)]

*Department of Physics, JRM Laboratory, Kansas State University, Manhattan, KS 66506, USA*

### Additional information for the experiment

The IR pump and the UV probe pulses were delayed with respect to each other by a mechanical variable delay stage and were spatially overlapped on the sample. The spatial overlap of the beams was ensured by steering them through a pinhole of 200 μm positioned in an equal distance to the sample outside of the chamber. All samples were obtained from commercial sources. The DWNTs were synthesized by the chemical vapor deposition (CVD) process, producing high purity nanotubes. The sample was attached to a Tantalum (Ta) substrate, and outgassed in multiple heating and annealing cycles with a peak temperature of 700 K. The sample temperature was measured with a type K thermocouple attached on the Ta disk. Ultra-high vacuum (UHV) of typically $10^{-10}$ mbar was maintained during these measurements.

---

[a)] Electronic mail: ichatzak@slac.stanford.edu

[b)] *Department of Materials Science and Engineering, Stanford University, Stanford, California 94305, USA and*

*Stanford Institute for Materials and Energy Sciences, SLAC National Accelerator Laboratory, Menlo Park, CA 94025, USA*

**Theoretical model**

In our model, the absorption is taken as instantaneous and the three-dimensional equations for the TTM can be reduced to two one-dimensional equations, since the laser spot size is much larger than the penetration depth. The source term $S(t)$ Eq. (1) has an exponential decay in space and a Gaussian profile in time, and describes the energy density, which excites the system. Due to the thickness of the sample (~100 μm bucky paper) used in this work and the transfer effect, the electron can penetrate unhindered into larger depths, resulting in a lower energy density in the excited area. This energy density is associated with the electron temperature. Hence, the latter shows a dependence on thickness[1]. Thus, we avoid the underestimation of the energy deposition depth by adding the ballistic range to the optical penetration depth in the source term, $S(x, t)$, of the TTM.[2]

$$S(x,t) = \sqrt{\frac{4ln2}{\pi}} \frac{(1-R)}{t_p(\delta + \delta_b)} F \cdot exp\left[-\frac{x}{(\delta + \delta_b)} - 4ln2\left[\left(\frac{t}{t_p}\right)^2\right]\right]$$

$$\times \left[\frac{1}{1-exp\left(-\frac{d}{\delta+\delta_b}\right)}\right], \quad (1)$$

where $R$ is the reflectivity, $F$ the total energy per pulse divided by the laser spot size, $t_p$ the full-width half-maximum (FHWM) of the excitation pulse, $\delta$ the penetration depth (~ 17nm), $\delta_b$ the ballistic range (~350 nm), and $d$ the sample thickness. The factor, $1/(1-exp(-d/\delta+\delta_b))$, has been incorporated into Eq. (1) to correct the finite thickness of the sample.

The boundary conditions Eq. (2) (Ref. 2 second part) used neglect losses from the front and back surfaces of the sample. In the model the sides of the sample had been constrained to the ambient temperature by adopting thermal-insulation boundary conditions. Thus, the losses at the front and back surfaces of the sample have been neglected. The electron and lattice temperatures were used as the two variables for which the two-coupled equations were solved.

$$\left.\frac{\partial T_e}{\partial x}\right|_{x=0} = \left.\frac{\partial T_e}{\partial x}\right|_{x=d} = \left.\frac{\partial T_l}{\partial x}\right|_{x=0} = \left.\frac{\partial T_l}{\partial x}\right|_{x=d} = 0$$

(2)

The initial conditions for the electron and the lattice systems were chosen as $T_e(x, -2t_p) =$

$T_l(x, -2t_p)$ = 300 K, where $t_p$ is the laser pulse FWHM. The two variables, which have been used for the solution of the differential equations, are the $T_e$ and $T_l$. We solved numerically the model of the two coupled equations using the finite element method.

The constant value of $2\times10^{16}$ W/m$^3$ for the coupling factor $G(T)$ applied in our calculations is a common method utilized in most of the theoretical investigations. However, there is experimental evidence suggesting the constant value may be applicable for experiments using low laser intensities and low electron temperatures.

The TR-TPP spectroscopy [3-5] used here enables us to study the relaxation dynamics of the charge carriers on a femtosecond timescale with very high resolution limited only by the resolution of the spectrometer.